\documentclass[aps,prc,superscriptaddress,showpacs,floatfix,nofootinbib,twocolumn]{revtex4-1}
\usepackage{amsmath,graphicx,float,hyperref}
\def\bra{\langle}
\def\ket{\rangle}
\newcommand{\trento}{T$\mathrel{\protect\raisebox{-2.1pt}{R}}$ENTo}

\begin{document}
\title{Nonlinear coupling of flow harmonics: Hexagonal flow and beyond}

\author{Giuliano Giacalone}
\affiliation{Institut de physique th\'eorique, Universit\'e Paris Saclay, CNRS, CEA, F-91191 Gif-sur-Yvette, France} 
\author{Li Yan}
\affiliation{Department of Physics, McGill University, 3600 University Street, Montreal, QC, H3A 2T8, Canada}
\author{Jean-Yves Ollitrault}
\affiliation{Institut de physique th\'eorique, Universit\'e Paris Saclay, CNRS, CEA, F-91191 Gif-sur-Yvette, France} 
%\date{\today}

\begin{abstract}
  Higher Fourier harmonics of anisotropic flow ($v_4$ and beyond) get large contributions induced by elliptic and triangular flow through nonlinear response.
  We present a general framework of nonlinear hydrodynamic response which encompasses the existing one, and allows to take into account the mutual correlation between the nonlinear couplings affecting Fourier harmonics of any order. 
  Using Large Hadron Collider data on Pb+Pb collisions at ~$\sqrt[]{s}=2.76$~TeV, we perform an application of our formalism to hexagonal flow, $v_6$, a coefficient affected by several nonlinear contributions which are of the same order of magnitude.
  We obtain the first experimental measure of the coefficient $\chi_{624}$, which couples $v_6$ to $v_2$ and $v_4$.
  This is achieved by putting together the information from several analyses: event-plane correlations, symmetric cumulants, as well as new higher-order moments recently analyzed by the ALICE collaboration. 
  The value of $\chi_{624}$ extracted from data is in fair agreement with hydrodynamic calculations, although with large error bars, which would be dramatically reduced by a dedicated analysis. 
    We argue that within our formalism the nonlinear structure of a given higher harmonic can be determined more accurately than the harmonic itself, and we emphasize potential applications to future measurements of $v_7$ and $v_8$.
   
\end{abstract}

\maketitle
\section{Introduction}

Anisotropic flow ($v_n$) in heavy-ion collisions~\cite{Heinz:2013th} has been measured up to the sixth Fourier harmonic, $v_6$~\cite{ATLAS:2012at,Chatrchyan:2013kba,Acharya:2017zfg}, and preliminary results on $v_7$ were recently reported~\cite{Tuo:2017ucz}. 
In ultra-central collisions, $v_n$ is to a good extent determined by linear response to the initial-state anisotropy in the harmonic $n$~\cite{Luzum:2012wu,CMS:2013bza}. 
In less central collisions, however, higher-order harmonics ($n\ge 4$) get important contributions induced by $v_2$ and $v_3$, through non-linear couplings~\cite{Gardim:2011xv}.
The magnitude of these non-linear couplings is encoded in the so-called \textit{response coefficients}, which are largely insensitive to the initial state and directly probe the hydrodynamic behavior~\cite{Yan:2015jma,Zhao:2017yhj}. 
As a consequence, a generic prediction of hydrodynamics is that these coefficients depend weakly on both the collision centrality and the details of hydrodynamic calculations~\cite{Borghini:2005kd,Teaney:2012ke,Qian:2016fpi}.
Therefore, nonlinear response coefficients are robust probes of hydrodynamic behavior: Any disagreement between the calculated values and experimental data cannot easily be fixed via a tuning of the parameters.

While it can be easily argued that there is only one leading nonlinear contribution to $v_4$ and $v_5$, several nonlinear couplings need to be considered in the decomposition of harmonics $v_6$~\cite{Bravina:2013ora} and higher.
Existing theoretical~\cite{Yan:2015jma,Qian:2016fpi,Zhao:2017yhj,Qian:2017ier,Liu:2018hjh} and experimental~\cite{Acharya:2017zfg} analyses of hexagonal flow isolate the various contributions by assuming that they are pairwise uncorrelated. 
For instance, they neglect the modest event-plane correlation between elliptic flow and triangular flow, which is measured~\cite{Aad:2014fla}.
In this article, we improve the existing formalism by relaxing this assumption. We show in Sec.~\ref{s:formalism} that even if the nonlinear terms are strongly correlated, they can still be separated by means of a simple matrix inversion. 
In Sec.~\ref{s:moments}, we explain how the corresponding matrix elements, which are moments~\cite{Bhalerao:2014xra}, can be obtained from existing data.
%on cumulants~\cite{Abelev:2014mda,Aaboud:2017acw,Sirunyan:2017fts}, event-plane correlations~\cite{Aad:2014fla}, symmetric cumulants~\cite{ALICE:2016kpq}, and higher-order moments which have recently been measured by the ALICE collaboration~\cite{Acharya:2017zfg}.
The values of the nonlinear response coefficients involving $v_6$ obtained from experimental data are presented in Sec.~\ref{s:results}, and they are compared to simple hydrodynamic calculations in Sec.~\ref{s:hydro}.
Eventually, in Sec.~\ref{s:new} we stress the importance of using our formalism in the characterization of the nonlinear structure of harmonics beyond hexagonal flow, $v_7$ and $v_8$. 
 
\section{Improved formalism of nonlinear coupling}
\label{s:formalism}

%add something more general about azimuthal transformation of vn and symmetry requirements on nonlinear response terms
%perhaps cite old papers having response that don't respect azimuthal symmetry, Petersen et al in 2010
Let us first recall how the nonlinear coupling is defined in the simple case of quadrangular flow, $v_4$~\cite{Yan:2015jma}.
In a hydrodynamic calculation, anisotropic flow is given in every event by $V_n\equiv\{ e^{in\varphi}\}$, where curly brackets denote an average value taken with the single-particle distribution at freeze-out~\cite{Cooper:1974mv,Teaney:2003kp}. 
The transformation of $V_n$ under an azimuthal rotation $\phi\to\phi+\alpha$ is $V_n\to V_ne^{in\alpha}$.
In this way, $V_4$ and $(V_2)^2$ both get the same factor $e^{4i\alpha}$, so that azimuthal symmetry allows a coupling between $V_4$ and $(V_2)^2$. 
Therefore, one can separate $V_4$ into a contribution proportional to $(V_2)^2$, and a remaining part, which we dub $U_4$\footnote{We always neglect contributions proportional to $V_1$, which is subleading.}:
\begin{equation}
\label{v4decomposition}
V_4=\chi_{42}(V_2)^2 +U_4,
\end{equation}
where the nonlinear response coefficient $\chi_{42}$ is the same for all events in a centrality class.
This decomposition is uniquely determined if one imposes the condition that the  two components $U_4$ and $(V_2)^2$ are uncorrelated, that is, $\langle U_4(V_2^*)^{2}\rangle=0$, where angular brackets denote an average over events in the centrality class.
This condition, together with Eq.~(\ref{v4decomposition}), implies
\begin{equation}
\label{defchi4}
\chi_{42}=\frac{\langle V_4(V_2^*)^2\rangle}{\langle |V_2|^4\rangle}. 
\end{equation}
This equation defines uniquely the response coefficient $\chi_{42}$, and was recently employed in experimental analyses~\cite{Acharya:2017zfg}. 
In an experiment, though, $V_n$ is not measured in every event due to finite multiplicity fluctuations, but the averaged quantities appearing in Eq.~(\ref{defchi4}) can be measured accurately \cite{Bhalerao:2014xra,Bhalerao:2011yg}.  
If $V_4$ is proportional to $(V_2)^2$ in every event, then $\chi_{42}$ as defined by Eq.~(\ref{defchi4}) reconstructs the proportionality coefficient, even if $V_2$ fluctuates event to event~\cite{Yan:2015jma}.
In this case, $U_4$ vanishes, so that $U_4$ can generally be interpreted as the part of $V_4$ which is not induced by $V_2$. 

Similarly, $V_5$ can be decomposed as
\begin{equation}
\label{v5decomposition}
V_5=\chi_{523}V_2V_3+U_5,
\end{equation}
hence
\begin{equation}
\label{defchi5}
\chi_{523}=\frac{\langle V_5V_2^*V_3^*\rangle}{\langle |V_2V_3|^2\rangle},
\end{equation}
which is also measured~\cite{Acharya:2017zfg}.

As for hexagonal flow, $V_6$, azimuthal symmetry allows several nonlinear terms~\cite{Qian:2016fpi}, so that it can be written as:
\begin{equation}
  \label{v6decomposition}
V_6=\chi_{62}(V_2)^3+\chi_{63}(V_3)^2+\chi_{624}V_2U_4+U_6.
\end{equation}
We write the third nonlinear term as $V_2U_4$ rather than $V_2V_4$ so as to avoid double counting with the $(V_2)^3$ term  in the case where $V_4$ is proportional to  $(V_2)^2$.
Note, however, that the structure of the decomposition is unchanged if one replaces $U_4$ with $V_4$.
Using Eq.~(\ref{v4decomposition}), one can indeed rewrite Eq.~(\ref{v6decomposition}) as:
\begin{equation}
 \label{v6decompositionbis}
V_6=(\chi_{62}-\chi_{624}\chi_{42})(V_2)^3+\chi_{63}(V_3)^2+\chi_{624}V_2V_4+U_6.
\end{equation}
%When written in this form, one can no longer assume that the nonlinear terms are mutually independent, as $V_4$ is strongly correlated with $(V_2)^2$. 

We show now that the nonlinear response coefficients appearing in Eq.~(\ref{v6decomposition}) are uniquely determined as soon as one imposes that all the nonlinear terms are uncorrelated with the last term, $U_6$.
In previous works~\cite{Yan:2015jma,Qian:2016fpi,Zhao:2017yhj,Qian:2017ier,Acharya:2017zfg}, though, such construction was supplemented by a stronger assumption, namely, that all  terms in the right-hand side of Eq.~(\ref{v6decomposition}) are pairwise independent. 
As we shall see, this turns out to be a reasonable approximation, but an unnecessary one. 

To simplify the notation, let us rewrite a decomposition such as (\ref{v6decompositionbis}), in the generic form
\begin{equation}
\label{decomp}
V=\sum_{k=1}^p\chi_k W_k+ U,
\end{equation}
where $p$ is the number of nonlinear terms ($p=3$ in the case of $V_6$), $W_k$ are products of lower-order harmonics ($W_1=(V_2)^3$, $W_2=(V_3)^2$, $W_3=V_2V_4$ for Eq.(\ref{v6decompositionbis})), and $\chi_k$ denote the corresponding coupling constants. 
As done for $V_4$, we define $U$ in Eq.~(\ref{decomp}) by the condition that it is linearly uncorrelated with all the nonlinear contributions:
\begin{equation}
\label{uncorrelated2}
\langle W_k^* U\rangle=0.
\end{equation}
This condition alone uniquely specifies the decomposition (\ref{decomp}). 
Multiplying Eq.~(\ref{decomp}) by $W_j^*$, averaging over events, and using Eq.~(\ref{uncorrelated2}), one obtains:
\begin{equation}
\label{system}
\langle W_j^*V\rangle=\sum_{k=1}^p \chi_k \langle W_j^*W_k\rangle.
\end{equation}
The left-hand side is a moment involving the higher harmonic $V$, while the terms $\langle W_j^*W_k\rangle$ are a set of moments involving lower-order harmonics. Note that since each $W_k$ is itself nonlinear, these moments are at least of order 4.
Eq.~(\ref{system}) is a linear system of $p$ equations for the $p$ coupling constants, $\chi_k$. 

We define a $p\times p$ matrix, $\Sigma$, by: 
\begin{equation}
\label{defsigma}
\Sigma_{jk}\equiv  \langle W_j^*W_k\rangle. 
\end{equation}
It is hermitian by construction, and real if one neglects parity violation~\cite{Fukushima:2008xe}.
We denote by $M$ the $p$-vector whose components are the moments $\langle W_j^*V\rangle$, and by $X$ the $p$-vector whose components are the response coefficients $\chi_k$. 
With these notations, the system (\ref{system}) can be rewritten in matrix form: 
\begin{equation}
\label{systemm}
M=\Sigma X.
\end{equation}
It is solved by inverting the matrix: 
\begin{equation}
\label{eq:chi}
X=\Sigma^{-1} M.  
\end{equation}
Assuming that all nonlinear terms are mutually independent, as done in previous works, amounts to assuming that $\Sigma$ is diagonal. 
As we shall see in Sec.~\ref{s:moments}, the off-diagonal elements of $\Sigma$ can all be extracted from existing data, hence, Eq.~(\ref{eq:chi}) can be applied directly without any approximation.

Note that these equations involves the higher harmonic $V$ only through the moments $\langle W_j^*V\rangle$ which are {\it linear\/} in the higher harmonic.
By constrast, standard measures of $v_n$, say, $v_n\{2\}\equiv \langle |V_n^2|\rangle$, are {\it quadratic\/}.
Since the magnitude of $v_n$ decreases rapidly with the order $n$, observables linear in a high-order harmonic $v_n$ are typically measured more accurately than quadratic observables. 
Hence, the nonlinear couplings of higher-order harmonics can be determined more precisely than these harmonics themselves~\cite{Acharya:2017zfg,Tuo:2017ucz}.

The quadratic mean $\langle |V^2|\rangle$ can be decomposed into the contributions of the various terms. 
Multiplying Eq.~(\ref{decomp}) by $V^*$, averaging over events, and using (\ref{uncorrelated2}), one obtains: 
\begin{equation}
\label{rmsV}
  \langle |V|^2\rangle=\sum_{k=1}^p \chi_k\langle V^*W_k\rangle+\langle |U|^2\rangle,
\end{equation}
which shows that each of the nonlinear couplings gives a separate contribution to $\langle |V^2|\rangle$ . 
%The uncertainty on the quadratic mean is typically dominated by the last term 

\section{Extracting the matrix elements from experimental data}
\label{s:moments}

We now explain how the nonlinear response coefficients for $V_6$ can be obtained from existing data on Pb+Pb collisions at $\sqrt{s}=2.76$~TeV.
If one decomposes $V_6$ according to Eq.~(\ref{v6decompositionbis}), the matrix (\ref{defsigma}) reads
\begin{equation}
\label{matrixv6}
\Sigma^{(6)} \equiv \Sigma = 
\begin{pmatrix}
\bra v_2^6 \ket  & \bra (V_2^*)^3 (V_3)^2 \ket & \bra v_2^2 V_4 (V_2^*)^2 \ket \cr
\bra (V_2)^3 (V_3^*)^2 \ket  & \bra v_3^4 \ket &\bra (V_3^*)^2 V_4 V_2 \ket  \cr
\bra v_2^2 V_4^* V_2^2 \ket  &  \bra V_3^2 V_4^* V_2^* \ket & \bra v_4^2v_2^2 \ket 
\end{pmatrix},
\end{equation}
where we have used the standard notation $v_n\equiv |V_n|$~\cite{Voloshin:1994mz}. 
The vector $M$ in Eq.~(\ref{eq:chi}) is
\begin{equation}
\label{vectorv6}
M=
\begin{pmatrix}
\bra (V_2^*)^3 V_6\ket\cr
\bra (V_3^*)^2 V_6\ket\cr
\bra V_2^*V_4^* V_6\ket
\end{pmatrix}.
\end{equation}
Note that the diagonal elements of $\Sigma$ are moments involving only the magnitude of anisotropic flow, $v_n$. 
On the other hand, the off-diagonal elements of $\Sigma$, as well as the components of $M$, involve relative phases between different Fourier harmonics, and are related to the so-called event-plane correlations~\cite{Bhalerao:2013ina,Aad:2014fla}. 

In principle, all these moments can be measured using the same experimental setup involving two subevents separated by a rapidity gap~\cite{Bhalerao:2014xra}. 
This method has recently been implemented by the ALICE collaboration~\cite{Acharya:2017zfg}. 
The values of $\Sigma_{11}$, $\Sigma_{22}$, $\Sigma_{13}$, $M_1$, $M_2$ can be directly obtained from these data through simple algebraic manipulations.\footnote{Specifically, in ALICE notation \cite{Acharya:2017zfg}, $\Sigma_{11}=(v_{6,222}/\chi_{6,222})^2$, $\Sigma_{22}=(v_{6,33}/\chi_{6,33})^2$, $M_1= v_{6,222}^2/\chi_{6,222}$,  $M_2= v_{6,33}^2/\chi_{6,33}$.} 
Note that $\Sigma_{13}= \bra v_2^2 V_4 (V_2^*)^2 \ket$ is a correlator  of higher order than that involved in the determination of the event-plane correlation between $V_2$ and $V_4$, namely, $\bra  V_4 (V_2^*)^2 \ket$. 
This higher-order correlator has been measured for the first time in Ref.~\cite{Acharya:2017zfg}.

For the remaining moments, we need to combine information from different analyses. $\Sigma_{12}$ could be extracted from the quantity dubbed $v^2_{3/\Psi_2}$ in the first ALICE analysis of triangular flow~\cite{ALICE:2011ab}. We instead choose to obtain it through the event-plane correlation measured by the ATLAS collaboration~\cite{Aad:2014fla} via
\begin{equation}
\langle\cos (6(\Phi_2-\Phi_3))\rangle_w=\frac{\Sigma_{12}}{\sqrt{\Sigma_{11}\Sigma_{22}}},
\end{equation}
where the left-hand side is in ATLAS notation. 
$\Sigma_{23}$ and $M_3$ are related to three-plane correlations through: 
\begin{eqnarray}
\label{3plane}
\langle\cos (2\Phi_2-6\Phi_3+4\Phi_4)\rangle_w&=&\frac{\Sigma_{23}}{v_2\{2\}v_3\{2\}v_4\{2\}},\cr
\langle\cos (2\Phi_2+4\Phi_4-6\Phi_6)\rangle_w &=&\frac{M_3}{v_2\{2\}v_4\{2\}v_6\{2\}}.
\end{eqnarray}
We extract the values of $\Sigma_{23}$ and $M_3$ from these equations using ATLAS data on event-plane correlations\footnote{In the centrality range where ATLAS uses 5\%  bins and ALICE uses 10\% bins, we take the average event-plane correlation in two consecutive bins in ATLAS data.}
and $v_n\{2\}$ from ALICE data~\cite{Acharya:2017zfg,Acharya:2017gsw}. 
It may not seem safe to mix data from two collaborations due to the different kinematic cuts. 
However, event-plane correlations should be largely independent of these cuts, as confirmed by the observation that ALICE and ATLAS values are compatible for those correlations, as reported in~\cite{Acharya:2017zfg}. 
%write something about the dominant uncertainty but this is probably too early, we should do it when we discuss the results. 
%the dominant uncertainty is on the last line, both the event)plane correlation and v62, but it should be much smaller if M3 was measured directly. 

Finally,  $\Sigma_{33}$ involves the correlation between the magnitude of different harmonics, $v_2$ and $v_4$. It is related to the so-called
symmetric cumulant $SC(4,2)$ recently measured by the ALICE collaboration~\cite{ALICE:2016kpq}:
\begin{equation}
\Sigma_{33}=SC(4,2)+v_2\{2\}^2v_4\{2\}^2.
\end{equation}

%are all the points included here? If some centralities are missing only for some elements, it should appear here. 
\begin{figure*}[t!]
\centering
\includegraphics[width=\linewidth]{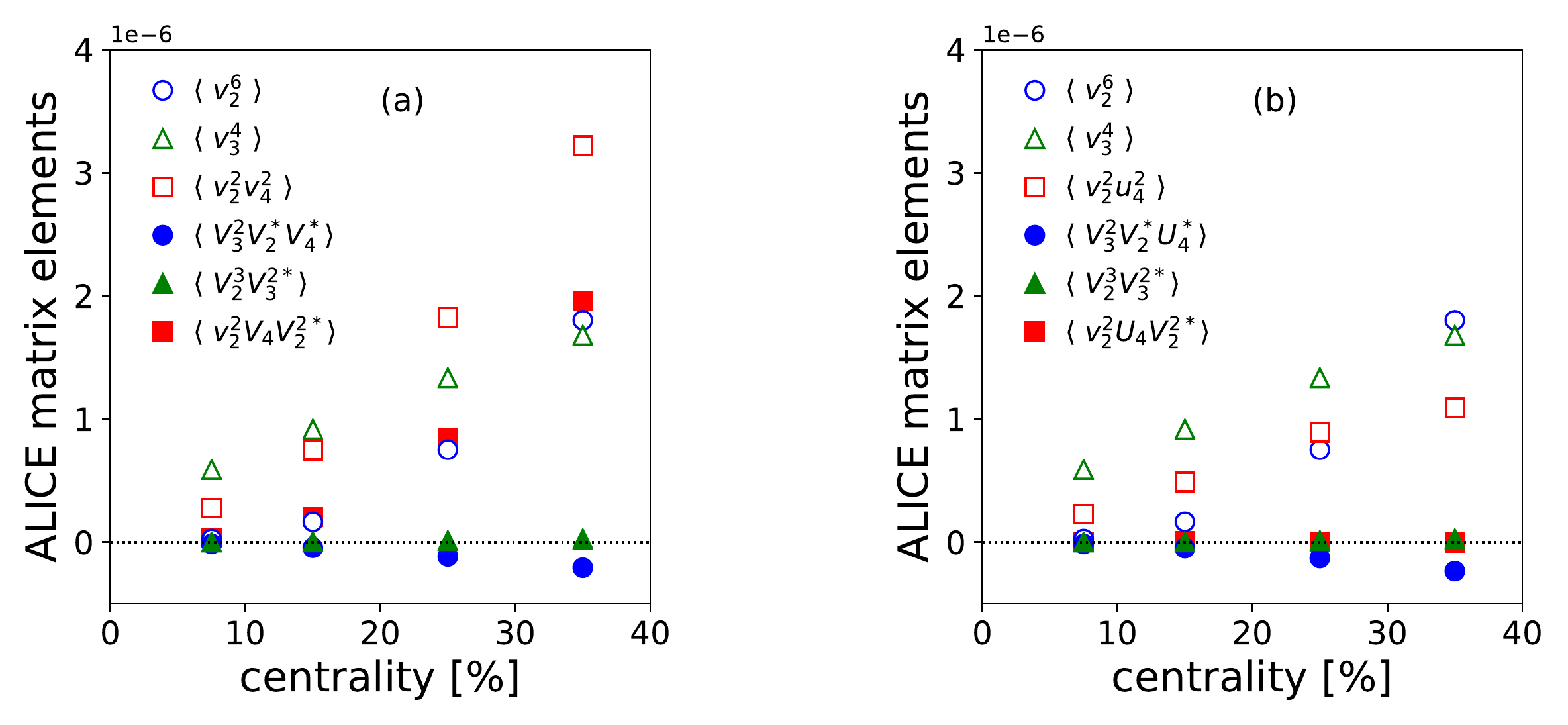}
\caption{(a) Elements of the matrix $\Sigma$ in Eq.~(\ref{matrixv6}) in Pb+Pb collisions at $\sqrt{s}=2.76$~TeV,
 as a function of centrality percentile. 
(b) Same matrix with $V_4$ replaced with $U_4$ everywhere (see text). We introduce the notation $u_4\equiv |U_4|$.}
\label{fig:1}
\end{figure*}
Figure~\ref{fig:1}--(a) displays the elements of $\Sigma$ as a function of the centrality percentile.\footnote{Since ALICE data on $\chi_{6,222}$ are available in the range 5-45\%, and on $\chi_{6,33}$ in the range 0-35\%, we are able to extract the matrix elements only in the range 5-35\%.} 
The off-diagonal element $\Sigma_{13}$ is large due to the strong correlation between $v_2$ and $v_4$. 
It is instructive to test how the matrix is modified when one writes the third nonlinear term in terms of 
$U_4$ (instead of $V_4$), as in Eq.~(\ref{v6decomposition}). 
Using Eq.~(\ref{v4decomposition}), one shows that this is done by transforming the matrix elements according to 
\begin{eqnarray}
\Sigma_{13}&\to&\Sigma_{13}-\chi_{42}\Sigma_{11},\cr
\Sigma_{23}&\to&\Sigma_{23}-\chi_{42}\Sigma_{21},\cr
\label{eq:19}\Sigma_{33}&\to&\Sigma_{33}-2\chi_{42}\Sigma_{31}+\chi_{42}^2\Sigma_{11}, 
%M_{3}&\to&M_{3}-\chi_{42}M_{1}.
\end{eqnarray}
where $\chi_{42}$ is measured by the ALICE collaboration~\cite{Acharya:2017zfg}. 
The transformed elements are displayed in Fig.~\ref{fig:1}--(b). 
When written in terms of $U_4$, the off-diagonal elements (full symbols) are much smaller than the diagonal elements (open symbols), which validates the approximations made in previous analyses~\cite{Acharya:2017zfg} where they were neglected. 
%not good: validate or not validate?
However, they are not compatible with zero. 
In particular, $\bra U_4 V_2 V_3^{2*} \ket$ is negative. 
It is therefore important to check to what extent the ALICE measurements, carried out under the assumption of negligible off-diagonal terms, are modified when the full pattern of correlations is taken into account. 
%Note that in this definition of $\Sigma^{(6)}$ we deal with a term $\bra u_4^2 v_2^2 \ket$.
%In the literature .
%Interestingly, ALICE data allow to test the validity of this assumption explicitly.
%We address this issue in Appendix~\ref{sec:A}

As a by-product of our analysis, we can test whether the two components in the decomposition of $V_4$, Eq.~(\ref{v4decomposition}), are independent.
We have imposed that they are uncorrelated, which is a weaker assumption.
If they are independent, it implies in addition that the matrix element $\bra v_2^2 u_4^2 \ket$ (open squares in Fig.\ref{fig:1}--(b)) factorizes into the product $\bra v_2^2 \ket \bra u_4^2 \ket$.
This can be directly tested using ALICE data, as discussed in Appendix~\ref{sec:A}.

\begin{figure*}[t!]
\centering
\includegraphics[width=\linewidth]{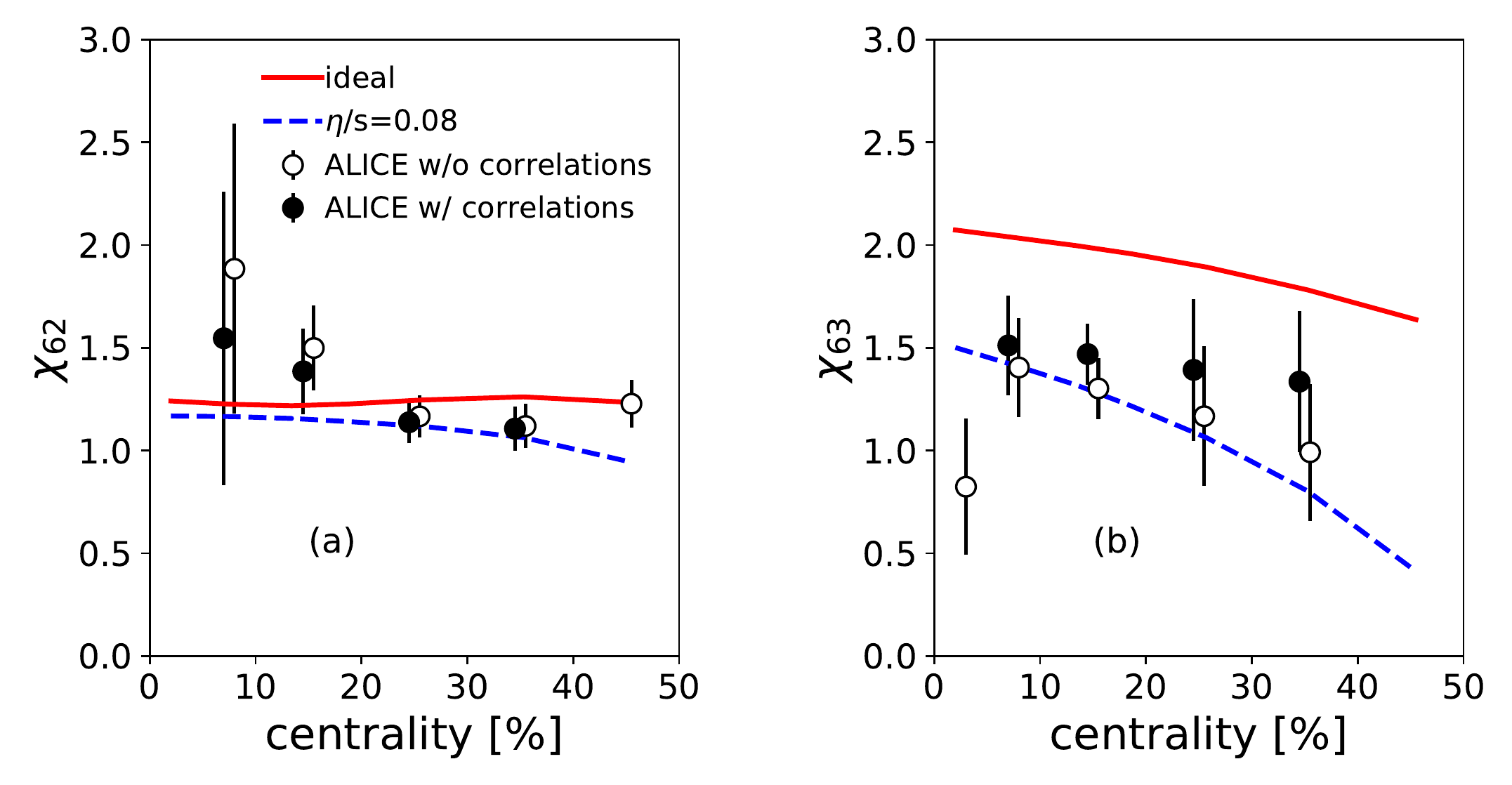}
\caption{Response coefficients $\chi_{62}$ (a) and $\chi_{63}$ (b) as a function of collision centrality in Pb+Pb collisions at $\sqrt{s}=2.76$~TeV.
  Open symbols: ALICE analysis~\cite{Acharya:2017zfg}. Full symbols: this analysis, which takes into account the mutual coupling between nonlinear terms.
We assume for simplicity that the error is unchanged. 
  Lines are hydrodynamic calculations (see Sec.~\ref{s:hydro}), ideal (solid) or viscous (dashed). 
}
\label{fig:2}
\end{figure*}
%\section{Hexagonal flow from nonlinear response}
\section{Nonlinear coefficients of $\boldsymbol{V_6}$ from data}
\label{s:results}

We now present our results for the nonlinear response coefficients of $v_6$ extracted from experimental data using   Eq.~(\ref{eq:chi}).  
The coefficient $\chi_{624}$ has already been calculated in hydrodynamics~\cite{Qian:2016pau,Zhao:2017yhj} but its experimental value is shown here for the first time. 
$\chi_{62}$ and $\chi_{63}$ have already been measured by ALICE under the approximation that the nonlinear terms are independent, while our new analysis takes into account the full correlation matrix. 

Our results for the coefficients  $\chi_{62}$ and $\chi_{63}$ are shown as full symbols in Fig.~\ref{fig:2}, and we compare them to the previous ALICE results, shown as open symbols.
Comparison between the two sets of points shows that mutual correlations between nonlinear terms in Eq.~(\ref{v6decomposition}) only have small effects. 
When they are taken into account, however, the centrality dependence of the response coefficient is somewhat flatter. As will be discussed in Sec.~\ref{s:hydro}, this generically improves agreement with hydrodynamics.
Taking into account the error bars, one cannot exclude that both response coefficients are independent of centrality. 
Note that, although within error bars, $\chi_{63}$ extracted from the full correlation matrix appears to be systematically larger that the previous ALICE data.
This is a signature of the largest non-diagonal term in $\Sigma$, namely, $\bra U_4 V_2 (V_3^*)^2 \ket$.
%Note that the error bar on $\chi_{62}$ is large in central collisions.
%The reason is that $v_2$ is small in central collisions, so that nonlinear effects associated with $v_2$ are much smaller. 

\begin{figure}[h]
\centering
\includegraphics[width=\linewidth]{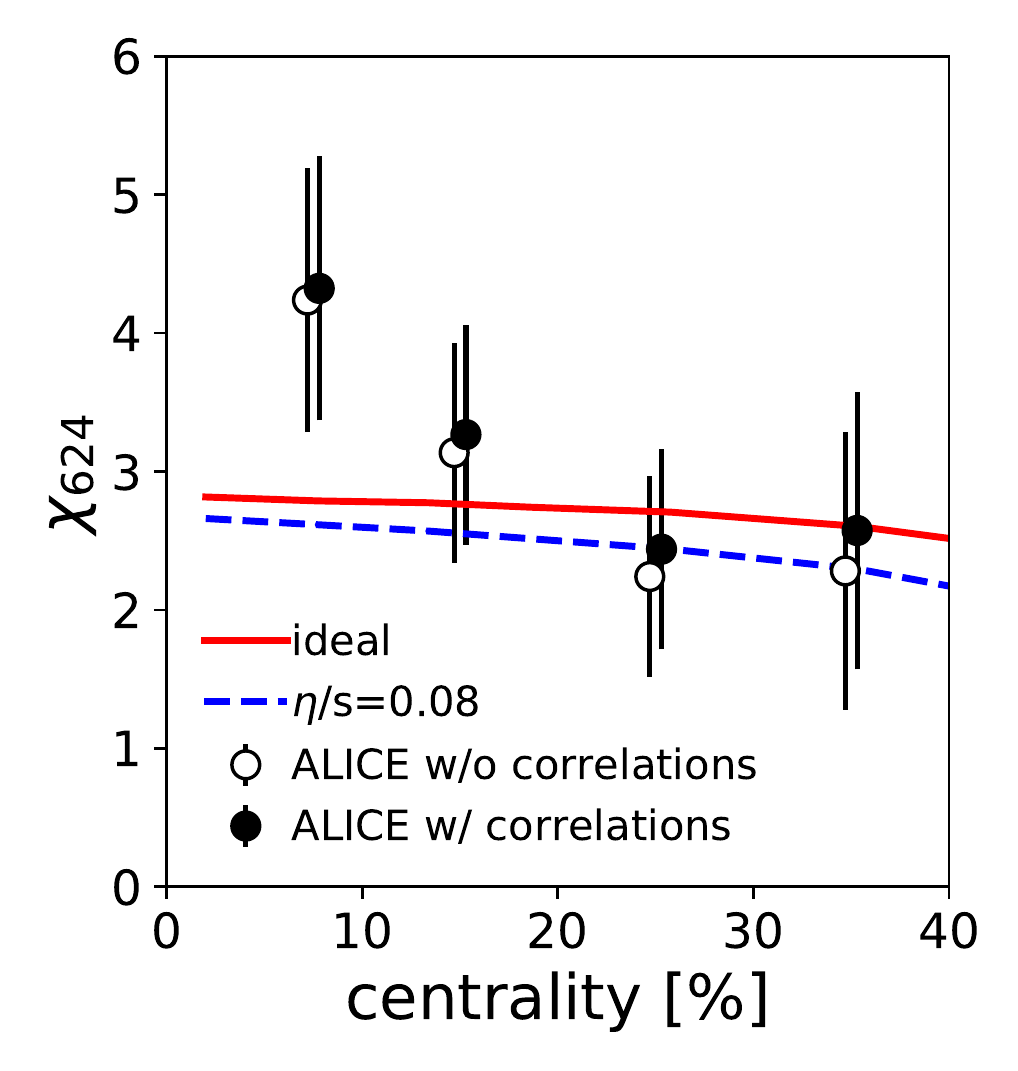}
\caption{Response coefficient $\chi_{624}$ as a function of collision centrality in Pb+Pb collisions at $\sqrt{s}=2.76$~TeV.
  Symbols denote the coefficient extracted from experimental data, both when the full correlation matrix is taken into account (open symbols), and when correlations are neglected (full symbols). 
Lines: Results from ideal (solid line) and viscous (dashed line) hydrodynamics.}
%\caption{Shaded band: $\chi_{624}$ coefficient extracted using Eq.~(5) and the elements of $\Sigma^{(6)}$. The error takes into account the uncertainty on $\bra \cos (2\Phi_2 + 4\Phi_4 - 6\Phi_6) \ket_w$ measured by the ATLAS Collaboration \cite{Aad:2014fla}. Solid line: Prediction of ideal hydrodynamics for $\chi_{624}$. Dashed line: Prediction of viscous hydrodynamics, with $\eta/s=0.08$. Dotted line: $\eta/s=0.16$}
\label{fig:3}
\end{figure}

Figure~\ref{fig:3} displays the first experimental result for $\chi_{624}$. 
It is larger than $\chi_{62}$ and $\chi_{63}$ for all centralities.
It also has a stronger centrality dependence, but the large errors prevent any definite conclusion. 
We estimate these errors by taking into account only the error on the event-plane correlation (second line of Eq.~(\ref{3plane})), which is the largest error.
If a dedicated analysis of $\chi_{624}$ was carried out, however, the error bar would likely be as small as that on $\chi_{62}$ and $\chi_{63}$.
The reason is that a dedicated analysis would measure directly $\langle V_2^*V_4^*V_6\rangle$, which is linear in $V_6$, while we extract it by combining the event-plane correlation and $v_6\{2\}$, which are both quadratic in $V_6$, and therefore have a larger error. 
%Likely, a direct analysis would return a value of $\chi_{624}$ with a flatter centrality dependence.
In Fig.~\ref{fig:3} (open symbols) we present as well the coefficient extracted from data in absence of mutual correlations between the nonlinear terms.
It is given by the following expression
\begin{equation}
\chi_{624}=\frac{\bra V_6 V_2^* U_4^* \ket}{\bra u_4^2 v_2^2 \ket},
\end{equation}
and it corresponds to the quantity computed in theoretical analyses \cite{Qian:2016fpi,Qian:2017ier,Zhao:2017yhj}.\footnote{These analyses further make the approximation $\bra u_4^2 v_2^2 \ket\approx\bra u_4^2 \ket \bra v_2^2 \ket$, whose validity is discussed in Appendix~\ref{sec:A}.}
This simplified expression returns a result very close to the full result. 
%As expected, we do not seize any visible difference between the fully correlated and the uncorrelated case.
%This means, essentially, that the assumptions so far made in the calculations of the nonlinear coefficients of hexagonal flow are justified by experimental data.

%This quantity has not been measured yet, and in hydrodynamics has been computed% assuming $\Sigma^{(6)}$ is diagonal, i.e., calculating
%\begin{equation}
%\chi_{624}' = \frac{\bra V_6 U_4^* V_2^* \ket}{\bra u_4^2 v_2^2\ket}.
%\end{equation}
%In Fig.~3 we show $\chi_{624}$ extracted using Eq.~(\ref{eq:chi}) and the matrix elements of Fig.~\ref{fig:1}.
%The shaded band represents the effect of including the error on $\bra \cos(2\Phi_2 + 4\Phi_4 - 6\Phi_6) \ket_w$ measured by the ATLAS Collaboration.
%The remarkable feature exhibited by this coefficient is its evident dependence on centrality.
%This is at variance with hydrodynamics calculation, where $\chi_{624}'$ is almost flat with centrality and decreases mildly if shear viscosity is implemented.
%We carry out such calculation also in this paper \textbf{* give details *}

%\section{Decomposition of $\boldsymbol{V_6}$}
\begin{figure}
\centering
\includegraphics[width=\linewidth]{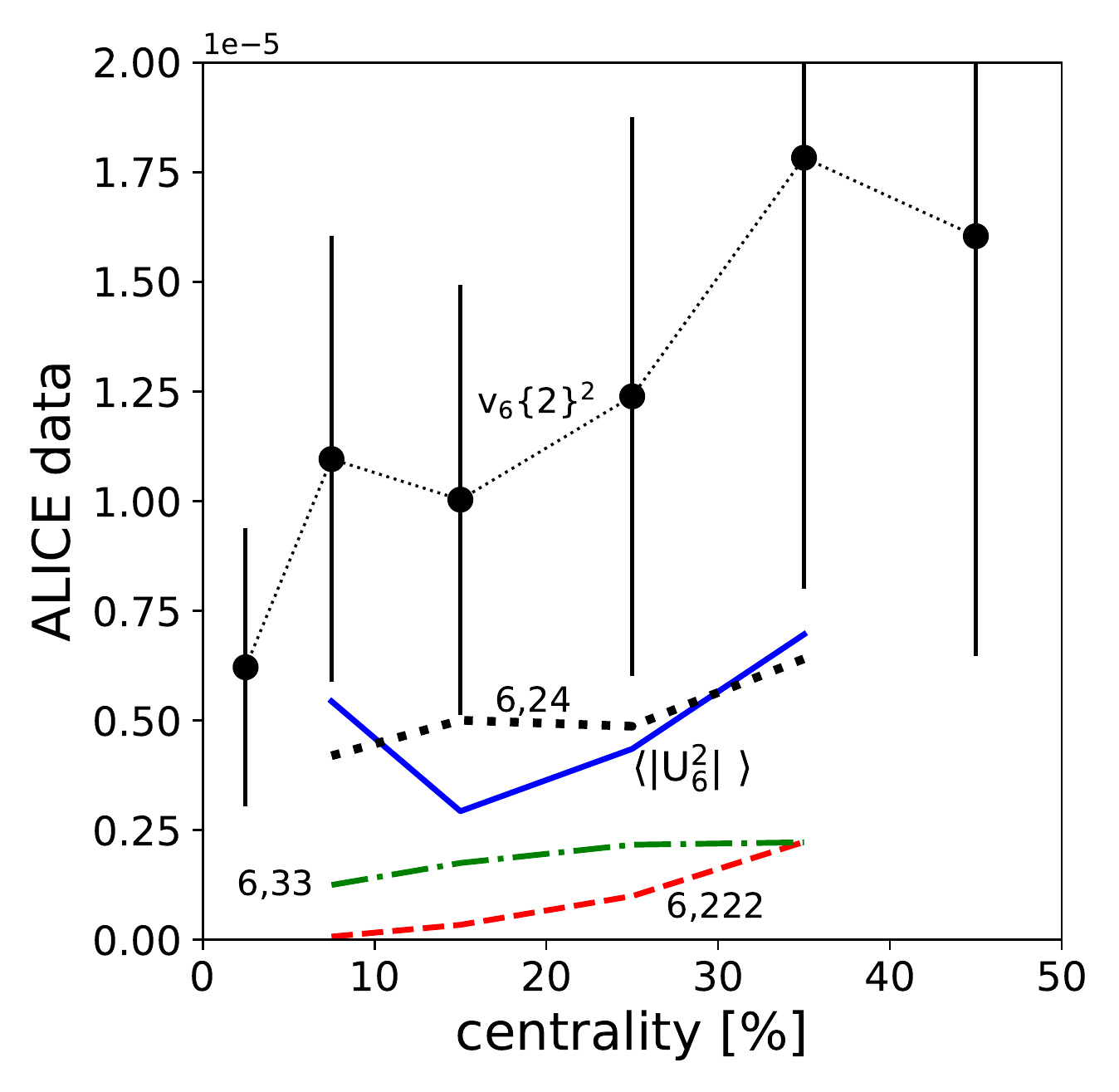}
\caption{Nonlinear contributions to $\bra|V_6|^2\ket$, as given by Eq.~(\ref{rmsv6}), extracted from ALICE data. Symbols: $\bra|V_6|^2\ket$ \cite{Acharya:2017zfg}. Solid line: $\bra |U_6|^2 \ket $. Thick dotted line: $\chi_{624}\bra V_6^* V_2 U_4 \ket$. Dot-dashed line: $\chi_{63}\bra V_6^* V_3^2 \ket$. Dashed line: $\chi_{624}\bra V_6^* V_2^3 \ket$.}
\label{fig:4}
\end{figure}

It is further instructive to compare the magnitudes of the various non-linear contributions to $V_6$.
In the case of the decomposition (\ref{v6decomposition}), Eq.~(\ref{rmsV}) reads:
\begin{equation}
\label{rmsv6}
  \bra|V_6|^2\ket=\bra|U_6|^2\ket+\chi_{62} \bra V_6^* V_2^3\ket+\chi_{63} \bra V_6^*V_3^2\ket+\chi_{624}\bra V_6^*V_2U_4\ket.
\end{equation}
Figure~\ref{fig:4} displays the values of all the terms appearing in this equation as a function of centrality.
$\bra|U_6|^2\ket$ is calculated as the difference between $v_6\{2\}^2$ and the sum of the other terms. %Therefore, we absorb into $u_6$ any potential subleading contribution neglected in our analysis, so that and the lines shown in the plot add exactly up to $v_6\{2\}^2$.
%We present for the first time, thus, the contribution to $v_6$ due to the coupling between $v_2$ and $v_4$. Interestingly, this turns out to be the largest contribution to the sixth flow harmonic.
The largest nonlinear contribution to $v_6$ is that due to the coupling with $v_2$ and $v_4$, while terms proportional to $(v_2)^3$ and $(v_3)^2$ are subleading for all centralities. 

\section{Hydrodynamic calculation}
\label{s:hydro}
Nonlinear response coefficients are unique probes of hydrodynamic behavior, because they are essentially independent of the initial state~\cite{Yan:2015jma}.
The uncertainty on the initial state is the bottleneck in traditional hydro-to-data comparisons~\cite{Luzum:2008cw}, because traditional flow observables such as $v_2$ or $v_3$ are driven by initial anisotropies in the corresponding harmonics, which are poorly constrained.
The dependence on these initial anisotropies cancels in the ratios defining the nonlinear response coefficients. 
This explains why all nonlinear response coefficients depend weakly on centrality in hydrodynamic calculations~\cite{Qian:2016fpi,Zhao:2017yhj}.
This is in sharp contrast with the steep centrality dependence of $v_2$.
Two different models of initial conditions with different $v_2$ and $v_3$ also return the same value for the nonlinear response coefficients.\footnote{Qian {\it et al\/}~\cite{Qian:2016fpi} report a significant difference between MC Glauber and MC KLN initial conditions, but only for 2 out of 8 coefficients, namely, $\chi_{42}$ and $\chi_{62}$. This difference is not seen by Zhao {\it et al\/}~\cite{Zhao:2017yhj}.}
%In hydrodynamic calculations, nonlinear response coefficients are mostly sensitive to the late dynamics. 
In this Section, we perform hydrodynamic calculations of the nonlinear response coefficients of $v_6$, and we compare them to the coefficients previously extracted from experimental data.

Because of the aforementioned weak dependence on initial state, we evaluate these coefficients in a single collision event with a smooth initial density profile.
The density profile is constructed by a smooth deformation of a symmetric 2-dimensional Gaussian, following Ref.~\cite{Teaney:2012ke}.
If one imprints a small elliptic deformation to the initial profile (a small asymmetry between the width along $x$ and $y$), the subsequent expansion generates all even Fourier harmonics, while odd harmonics (such as $V_3$) vanish because of $\phi\to\phi+\pi$ symmetry.
In this situation, $V_4$ is solely generated by $V_2$, so that $U_4$ vanishes.
Only the first nonlinear term remains in Eq.~(\ref{v6decomposition}), therefore, 
$\chi_{62}=V_6/(V_2)^3$. 
Similarly, one evaluates $\chi_{63}$ by imprinting a small triangular deformation to a radially symmetric profile, in which case  $\chi_{63}=V_6/(V_3)^2$. 
With this choice of initial conditions, the interpretation of $\chi_{62}$ and $\chi_{63}$ is transparent: they directly quantify the hexagonal flow produced by elliptic flow and triangular flow, respectively. 
%Li: what is the value of the deformation you chose? Can we be quantitative about these initial conditions? 

In the case of $\chi_{624}$, one needs to deform the initial profile in harmonics 2 and 4.
We carry out two hydrodynamic calculations, one with an asymmetric Gaussian profile, labeled (A) and one with a small quadrangular deformation (with the same orientation) on top of the asymmetric Gaussian, labeled (B). 
We compute $V_2$, $V_4$ and $V_6$ at the end of the hydrodynamic evolution for both initial conditions.
The initial conditions (A) and (B) differ only in the fourth Fourier harmonic, hence the values of $V_2$ are almost identical, $V_{2A}\simeq V_{2B}\equiv V_2$.
We first evaluate the change of $V_4$ induced by the quadrangular deformation, which is denoted by $U_4$ in Eq.~(\ref{v4decomposition}), and given by $U_4\equiv V_{4B}-V_{4A}$.
Then, the response coefficient $\chi_{624}$ defined by Eq.~(\ref{v6decomposition}) is given by the increase of $V_6$ driven by the quadrangular deformation, i.e.:
\begin{equation}
  \chi_{624}\equiv \frac{V_{6B}-V_{6A}}{V_{2}U_4}
  =\frac{V_{6B}-V_{6A}}{V_{2}(V_{4B}-V_{4A})}.
\end{equation}

We carry out this calculation for both ideal hydrodynamics and viscous hydrodynamics with $\eta/s=1/4\pi$~\cite{Kovtun:2004de}.
We compute pion spectra at a freeze-out temperature of $T_f=130$~MeV.
This simple setup is justified by previous studies which have shown~\cite{Qian:2016fpi} that a more elaborate calculation taking into account the full hadron spectrum and strong decays returns almost identical nonlinear response coefficients.
The resulting nonlinear response coefficients are plotted as lines in Figs.~\ref{fig:2} and \ref{fig:3}. 
Interestingly, the values for $\chi_{63}$ and $\chi_{624}$ are very close to those given by a full event-by-event hydrodynamic calculation~\cite{Qian:2016fpi}, implementing MC Glauber or MC KLN initial conditions.\footnote{For $\chi_{62}$, Qian {\it et. al}~\cite{Qian:2016fpi} find a different result depending on initial conditions. Their MC KLN result is similar to ours, while the MC Glauber result is much larger, and incompatible with data.}
Zhao {\it et al.\/}~\cite{Zhao:2017yhj} find significantly smaller values ($\chi_{63}$ is around 1, $\chi_{624}$ around 2) and do not comment on this difference with previous calculations.

In our hydrodynamic calculation, the centrality enters only through the initial transverse radius, which we estimate in a Glauber model. 
As the centrality percentile increases, this radius decreases and off-equilibrium effects become larger (earlier freeze out, and larger dissipative corrections both during the hydrodynamic expansion and at freeze out).
This explains the mild centrality dependence of nonlinear response coefficients.
This mild dependence is a characteristic of hydrodynamic models in general.
A steep variation of any nonlinear response coefficient would indicate a failure of the hydrodynamic picture.
Experimental results for $\chi_{62}$ and $\chi_{63}$ are so far compatible with hydrodynamics.
Our results on $\chi_{624}$, on the other hand, seem not capture the centrality dependence of the extracted experimental values.
This underlines the necessity of a dedicated analysis for reducing the error bars on this quantity.

Our results for $\chi_{62}$ and $\chi_{624}$ are mildly affected by adding shear viscosity to the hydrodynamic calculation. 
Viscosity results in a modest reduction of response coefficients, the effect being largest in peripheral collisions.
For $\chi_{63}$, we find a larger dependence on viscosity, and data are only compatible with viscous results. 
This large depenence is not observed  in event-by-event calculations~\cite{Qian:2016fpi}.

\section{Extension to higher harmonics}
\label{s:new}
The data-driven analysis carried out in the previous sections show that the assumptions made in the literature about $V_6$ are reasonable: The matrix $\Sigma^{(6)}$ is essentially diagonal.
In this section we argue, though, that it will be crucial to take into account the full pattern of correlations in harmonics of higher-order, such as $V_7$~\cite{Tuo:2017ucz}, and potentially $V_8$.

For heptagonal flow, there are also three leading nonlinear terms:
\begin{equation}
  \label{v7decomposition}
  V_7 = \chi_{723}V_2^2 V_3 + \chi_{725} V_2 U_5 + \chi_{734} V_3 U_4+U_7.
\end{equation}
Using Eqs.~(\ref{v4decomposition}) and (\ref{v5decomposition}), we rewrite the nonlinear terms as a function of the conventional harmonics $V_n$:
\begin{eqnarray}
  \label{v7decompositionbis}
  V_7 &=& (\chi_{723}-\chi_{725}\chi_{523}-\chi_{734}\chi_{42}) V_2^2 V_3\cr
  &&  + \chi_{725} V_2 V_5 + \chi_{734} V_3 V_4 +U_7,
\end{eqnarray}
which is again of the type (\ref{decomp}) with $W_1=V_2^2 V_3$,
$W_2=V_2V_5$, $W_3=V_3V_4$.
The correlation matrix (\ref{defsigma}) is: 
\begin{equation}
  \label{v7matrix}
\Sigma^{(7)} = 
\begin{pmatrix}
\bra v_2^4 v_3^2 \ket  & \bra v_2^2 V_2^* V_3^* V_5 \ket & \bra v_3^2 V_2^{2*} V_4 \ket  \\
 \bra v_2^2 V_2 V_3 V_5^* \ket  & \bra v_2^2 v_5^2 \ket &\bra V_5^* V_2^* V_3 V_4 \ket  \\
 \bra v_3^2 V_2^2 V_4^* \ket  &  \bra V_5 V_2 V_3^* V_4^* \ket & \bra v_3^2 v_4^2 \ket 
\end{pmatrix}.
\end{equation}
Note that $\Sigma_{23}$ involves four different Fourier harmonics, and is related to a 4-plane correlation~\cite{Bhalerao:2013ina}. 
Hydrodynamic calculations of $\chi$ coefficients involving $V_7$ are already on the market \cite{Qian:2016fpi,Qian:2017ier,Zhao:2017yhj}, and always implicitly assume that the matrix is diagonal.
It will be interesting to see if the hydrodynamic results for $\chi_{723}$, $\chi_{725}$ and $\chi_{734}$ are modified once the full correlation structure is taken into account.

For completeness, let us also provide the correlation matrix of $V_8$, a coefficient which is likely to be accessible to experimental analyses thank to the massive statistics of data collected in Pb+Pb collisions at LHC2.
We decompose $V_8$ as follows: 
\begin{align}
\nonumber V_8 = \chi_{82}V_2^4 +& \chi_{823}V_2V_3^2 + \chi_{824}V_2^2U_4 + \chi_{826} V_2U_6\\ 
&+\chi_{835} V_3U_5 + \chi_{84} U_4^2 + U_8,
\end{align}
which, in terms of the harmonics $V_n$, reads
\begin{align}
\nonumber V_8 = &\bigl( \chi_{82} - \chi_{824}\chi_{42}-\chi_{826}\chi_{62}+ \\
\nonumber &\hspace{25pt}+ \chi_{826}\chi_{624}\chi_{42} + \chi_{84}\chi_{42}^2 \bigr)V_2^4 \\
\nonumber &+\bigl( \chi_{823}-\chi_{826}\chi_{63}-\chi_{835}\chi_{523}\bigr)V_2V_3^2 \\
\nonumber &+\bigl( \chi_{824} - \chi_{826}\chi_{624}-2\chi_{84}\chi_{42}\bigr)V_2^2V_4\\
&+\chi_{826}V_2V_6 + \chi_{835}V_3V_5 + \chi_{84}V_4^2+U_8.
\end{align}
This leads to the following $6\times 6$ correlation matrix:
\begin{widetext}
\begin{equation}
  \label{v8matrix}
\Sigma^{(8)} = 
\begin{pmatrix}
\bra v_2^8 \ket  & \bra v_2^2 V_3^2 V_2^{3*} \ket & \bra v_2^4 V_4 V_2^{2*} \ket & \bra v_2^2 V_6 V_2^{3*} \ket & \bra V_5V_3V_2^{4*} \ket & \bra V_4^2 V_2^{4*} \ket \\
{\rm c.c.}  & \bra v_2^2 v_3^4 \ket & \bra v_2^2 V_2 V_4 V_3^{2*} \ket & \bra v_2^2 V_6 V_3^{2*} \ket & \bra v_3^2V_5V_2^*V_3^* \ket & \bra V_4^2 V_2^*V_3^{2*} \ket \\
{\rm c.c.}  & {\rm c.c.} & \bra v_2^4v_4^2 \ket & \bra v_2^2 V_6 V_2^* V_4^* \ket & \bra V_5V_3V_2^{2*}V_4^* \ket & \bra v_4^2 V_4 V_2^{2*} \ket \\
{\rm c.c.}  & {\rm c.c.} & {\rm c.c.} & \bra v_2^2v_6^2 \ket & \bra V_5V_3V_2^*V_6^* \ket & \bra V_4^2 V_2^*V_6^* \ket \\
{\rm c.c.}  & {\rm c.c.} & {\rm c.c.} & {\rm c.c.} & \bra v_5^2v_3^2 \ket & \bra V_4^2 V_3^*V_5^* \ket \\
{\rm c.c.}  & {\rm c.c.} & {\rm c.c.} & {\rm c.c.} & {\rm c.c.} & \bra v_4^4 \ket
\end{pmatrix}.
\end{equation}
\end{widetext}
%Even in the basis of $U_n$, it is not reasonable to assume that the matrix is diagonal, and it is mandatory to take into account all possible mutual correlations.
%JYO: I am commenting this out as there is no argument, you are just stating that it is not reasonable. 
Let us stress once more, then, that the extraction of the nonlinear coefficients in our framework involves only moments which are linear in $V_8$, and, therefore, experimentally easier to achieve than typical observables such as $\bra v_8^2 \ket$.

\section{Conclusion}
We have proposed a new framework which allows to systematically isolate the various nonlinear contributions to a given higher-order harmonic, and measure the nonlinear coupling coefficients. 
It can be applied to both experimental data and event-by-event hydrodynamic calculations. 
%The improved framework is completely generic and allows to introduce more and more terms in the decomposition of any flow harmonic in a systematic way. 
The main improvement over previous analyses is that we take into account the mutual correlations between nonlinear contributions.
We have applied this new framework to $V_6$ using existing data. 
When mutual correlations are properly taken into account, the centrality dependence of the response coefficients $\chi_{62}$ and $\chi_{63}$ becomes somewhat flatter, thus improving agreement with hydrodynamic predictions. 
%Our analysis indicates that such correlations are very mild in the case of hexagonal flow, as they do not affect the nonlinear response coefficients involving $v_6$.
We have provided the first experimental determination of the coefficient $\chi_{624}$ coupling $v_6$ to $v_2$ and $v_4$. 
It is in fair agreement with hydrodynamic predictions, though with large error bars. 
The corresponding nonlinear term is the largest of the three nonlinear contributions to $v_6$. 
%Further, experimental data indicate that the largest contribution to $v_6$ comes indeed from the coupling between $v_2$ and $v_4$.
With the advent of large statistics Pb+Pb LHC2 data, we expect that this new framework will enable detailed analyses of higher-order flow coefficients, which will provide precision tests of hydrodynamic behavior. 
%the presented framework to become the standard paradigm in the analysis of anisotropic flow coefficients of higher order.

%\section{Acknowledgements}

\appendix
\section{Correlation between $\boldsymbol{u_4^2}$ and $\boldsymbol{v_2^2}$}
\label{sec:A}
In the definition of quadrangular flow in Eq.~(\ref{v4decomposition}), one chooses $U_4$ and $V_2^2$ to be uncorrelated by construction, i.e.,
\begin{equation}
\label{eq:uncorrel}
\bra U_4^* V_2^2 \ket = 0.
\end{equation}
In the literature~\cite{Qian:2016fpi,Giacalone:2016afq,Qian:2017ier,Zhao:2017yhj}, though, this has always been supplemented by the following assumption:
\begin{equation}
\label{equal} \bra u_4^2 v_2^2 \ket =\bra u_4^2 \ket  \bra v_2^2 \ket,
\end{equation}
which implies statistical independence between the two terms.
Statistical independence is a much stronger constraint than just requiring the two terms to be uncorrelated.

\begin{figure}[h]
\centering
\includegraphics[width=\linewidth]{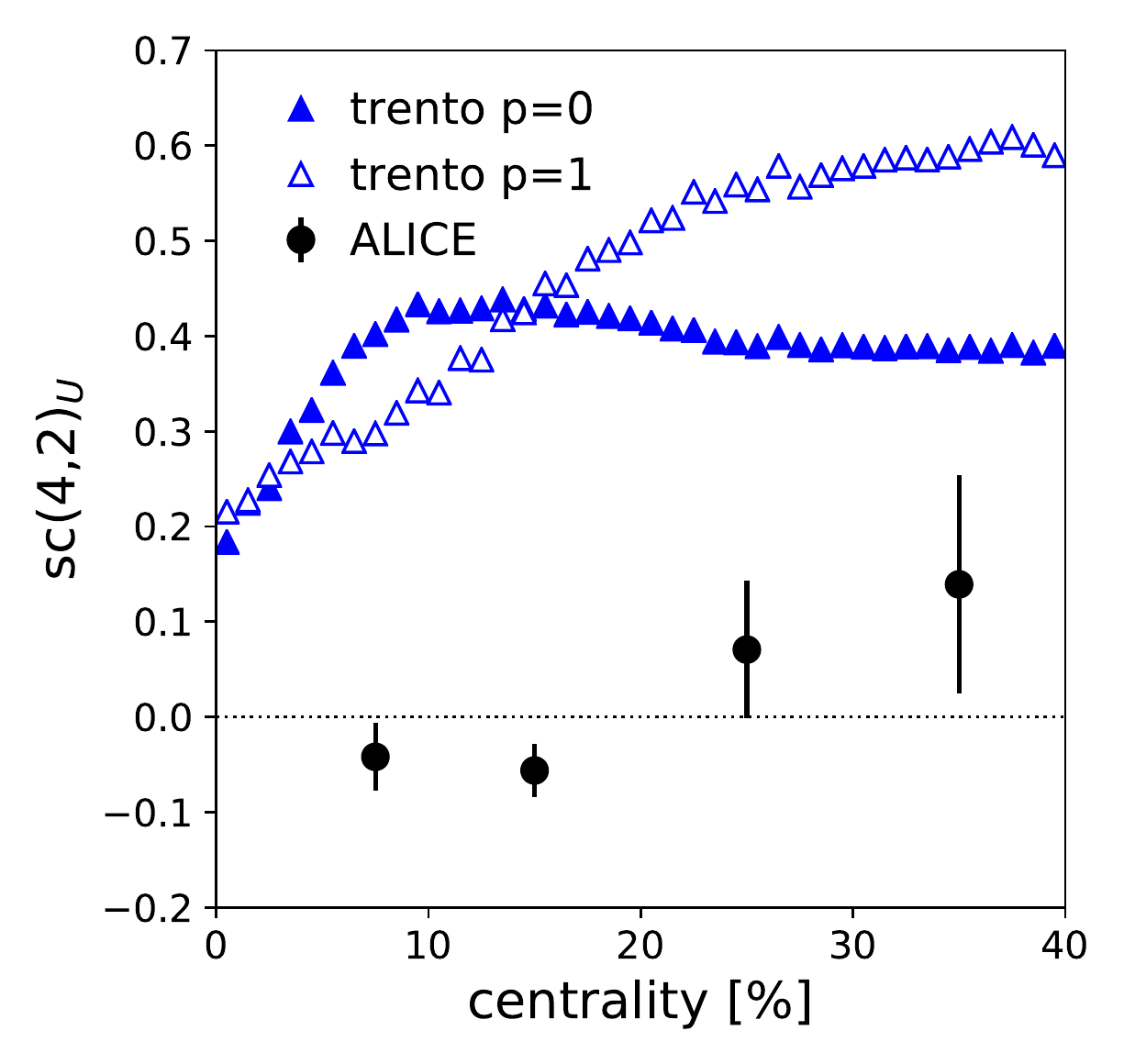}
\caption{Circles: $sc(4,2)_U$, as given by Eq.~(\ref{eq:scU}), from ALICE data. Triangles: \trento{} calculations for $sc(4,2)_\varepsilon$, Eq.~(\ref{sc42uc}).}
\label{fig:5}
\end{figure}

In this appendix we check the validity of Eq.~(\ref{equal}) using experimental data.
The equality can be conveniently tested by introducing the following normalized symmetric cumulant~\cite{Bilandzic:2013kga,ALICE:2016kpq}
\begin{equation}
\label{eq:scU}
sc(4,2)_U = \frac{\bra u_4^2 v_2^2\ket- \bra u_4^2 \ket \bra v_2^2\ket}{\bra u_4^2 \ket \bra v_2^2\ket},
\end{equation}
which vanishes if $u_4^2$ and $v_2^2$ are independent.
All the terms appearing in the definition of $sc(4,2)_U$ are available from ALICE data.
The quantity $\bra u_4^2 \ket$ was recently measured~\cite{Acharya:2017zfg}, and $\bra u_4^2 v_2^2 \ket$ is given by $\Sigma_{33}$ in Eq.~(\ref{eq:19}).
The resulting cumulant is displayed as black circles in Fig.~\ref{fig:5}.
This result illustrates that $sc(4,2)_U$ is indeed small in magnitude, so that Eq.~(\ref{equal}) is a good approximation, although error bars, driven by the uncertainty on SC(4,2), are large in non-central collisions.

We ask now whether this quantity carries any useful information about the initial state of the hydrodynamic evolution.
In heavy-ion collisions, elliptic flow is to a good approximation proportional to the second eccentricity harmonic of the initial state, $\varepsilon_2$ \cite{Teaney:2010vd}.
As for the fourth harmonic, it is often argued that $U_4$ may scale linearly with the fourth cumulant eccentricity of the initial medium \cite{Qian:2017ier}, which is customarily taken from Ref.~\cite{Teaney:2012ke}:
\begin{equation}
\nonumber \mathcal{C}_4 = \mathcal{E}_4 + 3 \frac{\bra r^2\ket^2}{\bra r\ket^4} \mathcal{E}_2^2, 
\end{equation}
where $\mathcal{E}_n$ corresponds to the moment-defined eccentricity harmonic of order $n$ \cite{Teaney:2010vd}.

Now, dubbing $c_4=|\mathcal{C}_4|$, if $\mathcal{C}_{4}$ scales linearly with $U_4$, then the following quantity
\begin{equation}
\label{sc42uc}
sc(4,2)_\varepsilon=\frac{\bra c_4^2 \varepsilon_2^2 \ket - \bra c_4^2 \ket \bra \varepsilon_2^2 \ket}{\bra c_4^2 \ket \bra \varepsilon_2^2 \ket}
\end{equation}
should match to a good extent $sc(4,2)_U$ observed in ALICE data, at least in central collisions where $U_4$ dominates.
To check this, we compute $sc(4,2)_\varepsilon$ from initial-state  simulations of Pb+Pb collisions at $~\sqrt[]{s}=2.76$~TeV.
We perform this by employing two different setups of the \trento{} model of initial conditions \cite{Moreland:2014oya}.
We use \trento{} with $p=1$, corresponding to a Glauber Monte Carlo model, and $p=0$, which provides eccentricities in agreement with models including high-energy QCD effects, such as EKRT or IP-Glasma~\cite{Moreland:2014oya}.
Our results from the initial state models are shown in Fig.~\ref{fig:5}.
The correlation is moderate and positive already at 0\% centrality, therefore, we do not find agreement between the model calculation and experimental data.
Therefore, typical models of initial conditions do not suggest a linear correlation between $\mathcal{C}_4$ and $U_4$, in contradiction, then, with the finding of hydrodynamic calculations~\cite{Qian:2017ier}.
We remark, though, that this failure may be equally due to either the fact that the fourth-order initial-state anisotropy is not correctly quantified by $\mathcal{C}_4$, or that $\mathcal{C}_4$ is not captured by the initial-state models, or both.

\end{document}